\pdfoutput=1
\documentclass[runningheads]{llncs}
\usepackage[utf8]{inputenc}
\usepackage{hyperref}
\usepackage{subcaption}
\usepackage{color}
\usepackage{color, colortbl}
\usepackage[usenames,dvipsnames]{xcolor}
\usepackage{multirow}
\usepackage{booktabs}
\usepackage[font=small,labelfont=bf]{caption}
\usepackage{adjustbox}

\usepackage{ amssymb }

\usepackage{graphicx}
\usepackage{tikz}
\usepackage{cite}
\usepackage{lipsum}
\usepackage[inline]{enumitem}
\newcommand{\limit}{\textbf{exploration}}
\newcommand{\Limit}{\textbf{Exploration}}
\newcommand{\ex}{$\xi$}
\newcommand{\lam}{\textbf{learning speed}}
\newcommand{\Lam}{\textbf{Learning Speed}}
\newcommand{\ls}{$\lambda$}
\newcommand{\threshold}{\textbf{tolerance}}
\newcommand{\Threshold}{\textbf{Tolerance}}
\newcommand{\tl}{$\tau$}
\newcommand{\subtopic}{\textbf{subtopic switching}}
\newcommand{\Subtopic}{\textbf{Subtopic Switching}}
\newcommand{\sub}{$\varphi$}

\newcommand{\greedy}{\textbf{Greedy}}
\newcommand{\random}{\textbf{Random}}
\newcommand{\reverse}{\textbf{Reverse}}
\newcommand{\greedys}{\textbf{Greedy-Skip}}

\newcommand{\simiir}{\texttt{SimIIR}}

\newcommand{\progress}{\texttt{FEEDBACK$_{\texttt{SC}}$}}
\newcommand{\car}{TREC CAR}
\newcommand{\sal}{SAL}
\newcommand{\sacsm}{\texttt{SACSM}}
\newcommand{\csm}{\texttt{CSM}}

\definecolor{modelblue}{rgb}{0.0, 0.45, 0.78}

\newcommand*\circled[1]{\tikz[baseline=(char.base)]{\node[shape=circle,fill=black,inner sep=1pt] (char) {\textcolor{white}{#1}};}}

\newcommand\blfootnote[1]{%
  \begingroup
  \renewcommand\thefootnote{}\footnote{#1}%
  \addtocounter{footnote}{-1}%
  \endgroup
}

\begin{document}
    
\title{Searching, Learning, and Subtopic Ordering:\\
A Simulation-based Analysis}

\author{Arthur C\^{a}mara \and David Maxwell \and Claudia Hauff}

\authorrunning{A. C\^{a}mara, D. Maxwell and C. Hauff}
\titlerunning{SAL and Subtopic Ordering: A Simulation-based Analysis}

\institute{Delft University of Technology\\Delft, The Netherlands\\
\email{\{a.barbosacamara,d.m.maxwell,c.hauff\}@tudelft.nl}}

\maketitle
  
\begin{abstract}

Complex search tasks---such as those from the \textit{Search as Learning (SAL)} domain---often result in users developing an information need composed of several aspects. However, current models of searcher behaviour assume that individuals have an \textit{atomic} need, regardless of the task.
While these models generally work well for simpler informational needs, we argue that searcher models need to be developed further to allow for the decomposition of a complex search task into multiple aspects. 
As no searcher model yet exists that considers both aspects and the SAL domain, we propose, by augmenting the \textit{Complex Searcher Model} (\csm{}), the \textit{Subtopic Aware Complex Searcher Model} (\sacsm{})---modelling aspects as subtopics to the user's need. We then instantiate several agents (i.e., simulated users), with different \textit{subtopic selection} strategies, which can be considered as different prototypical learning strategies (e.g., \textit{should I deeply examine one subtopic at a time, or shallowly cover several subtopics?}). Finally, we report on the first large-scale simulated analysis of user behaviours in the SAL domain.
Results demonstrate that the \sacsm{}, under certain conditions, simulates user behaviours accurately.
\end{abstract}

\section{Introduction}\label{sec:intro}


Over the years, a series of models\footnote{In this paper, we refer to a \textit{model} as a \textit{\textbf{model of user behaviour}}.} that describe searcher behaviour have been defined~\cite{bates1989design,belkin1990cognitive,kuhlthau1988developing}. These often provide a post-hoc explanation of---and reasoning behind---the actions of a searcher during information seeking. One of the main drawbacks of such models is their lack of predictive capabilities: we can neither use these models to investigate what is likely to occur in different instantiations of a retrieval system; nor can we use them for simulating user behaviour.\blfootnote{This research has been supported by \textit{NWO} projects \textit{SearchX} (\texttt{639.022.722}) and \textit{Aspasia} (\texttt{015.013.027}).}

Indeed, models examining searcher behaviours with predictive power~\cite{azzopardi2011economics,azzopardi2014modelling,fuhr2008probability} have only recently been explored in the field of \textit{Interactive Information Retrieval (IIR)}. Such models enable us to relate changing costs (e.g., the cost of examining a document) to changing searcher behaviours.
Prior works employing these models have investigated how searchers interact with ranked lists~\cite{moffat2012models}, the impact of different browsing costs on a searcher's behaviour~\cite{azzopardi2016two,kashyap2010facetor}, and stopping behaviours on \textit{Search Engine Results Pages (SERPs)}~\cite{maxwell2018information,wu2014using}. 
Search topics are usually considered \textit{atomic} in all of these prior works, with a simple information need. That is, over a \textit{search session}\footnote{We consider a \textit{search session} as interactions with a search interface, which can include the issuing of multiple queries---and the examination of multiple documents.}, a single topic is considered---with retrieved documents considered to be either relevant or non-relevant to that one topic. 
These works do not consider the different \textit{aspects} that may constitute a wider topic. In this work, we introduce the first model of user behaviour that incorporates such thinking. More specifically, we take as a starting point the \textit{Complex Searcher Model (\csm{})}~\cite{maxwell2016simulating}, a model that considers a user's interactions throughout a search session (over multiple queries), and extend it to yield the \textit{Subtopic-Aware Complex Searcher Model (\sacsm)}---which, by considering the aspects as subtopics of a larger information need, models: \textit{(i)} subtopic selection; and \textit{(ii)} subtopic switching steps in the search process.

With the \sacsm{}, we explore the effect of different subtopic switching strategies for multiple types of users within a particular domain to ground our work. We consider \emph{Search as Learning (SAL)}, defined by Marchionini~\cite{marchioni06exploratory}, as an iterative process whereby learners engage by reading, scanning and processing a large number of documents retrieved by a search system. Here, the goal is to gain knowledge about a specific learning objective. With web search engines having become an essential resource for learners~\cite{eickhoff2014lessons}, it is therefore vital to provide support to learners (e.g., through the form of novel interface designs~\cite{roy2021note, camara2021searching, roy2021incorporating} or rankers optimised for human learning~\cite{syed2017retrieval}) that help improve their learning efficiency while searching.
As a learner's complex information needs can often be decomposed into several subtopics, a natural question to ask is \textit{how searchers should tackle the different subtopics to learn efficiently.}

To answer this question, we present an exploratory study of the \sacsm{} where we simulate different types of learners as \textit{agents}\footnote{
\textit{Agents} are \textit{simulated users} that are able to make judgements as to the relevancy/attractiveness of information \textit{without} recourse to relevance information~\cite{maxwell16agents}.}, and compare these to each other, examining the effect their search behaviour has on their ability to discover documents containing important keywords, as well as how they navigate throughout the subtopic space. 
We instantiate a series of agents that subscribe to the \sacsm{}---with four tunable parameters that control their simulated searching behaviour: \textit{(i)} \lam{} (\ls), or how fast agents incorporate novel terms into their vocabulary; \textit{(ii)} \limit{} (\ex), or how willing agents are to explore each subtopic; \textit{(iii)} \threshold{} (\tl), or how willing an agent is to click on a search result snippet; and \textit{(iv)} ~\subtopic{} (\sub), the strategy that agents employ to navigate through subtopics. As such, we present the first SAL study that employs simulation to examine the search behaviours of learners. By grounding a series of simulated agents with interaction data from a prior user study, we run extensive simulations of interaction to address the following research question:
\begin{enumerate}
    \item[\textbf{RQ}]{\textit{How do \subtopic{} (\sub{}) strategies for learning-oriented search tasks affect the search behaviour of simulated agents?}}
\end{enumerate}

To answer \textbf{RQ}, we measure behaviours by tracking how specific measures---the \emph{number of keywords found}, the \emph{order of keywords found}, and \emph{subtopic exploration}---evolve over an agent's search sessions. We argue that to be considered effective, a strategy should allow an agent to: \textit{(i)} discover as many keywords as possible in the early stages of the session; and \textit{(ii)} help the agent to complete the subtopic space exploration in as few steps as possible.

The main findings of our work are: \textit{(i)} subtopic switching strategies that prioritise ordering in the subtopic picking process yield improved discovery of keywords and exploration of subtopics; and \textit{(ii)} the \sacsm{} is enough to instantiate agents that display behaviour similar to real-world learners in a SAL context. Findings suggest that the \sacsm{} is a high-quality model that provides a solid step in approximating searcher behaviours in the SAL domain. This is vital for works that rely on large-scale simulations, such as reinforcement learning for training new rankers optimised for human learning, as well as quickly evaluating new interfaces and algorithmic changes cheaply---all in a simulated environment.
\section{Related Work}\label{sec:background}

\paragraph{\textbf{Models of Searcher Behaviour}} Models of searcher behaviour typically fall into one of two categories: \textit{(i)} descriptive models \cite{bates1989design,pirolli1999information,kuhlthau1988developing,ellis1993modeling,ingwersen2006turn}, allowing us to gain an intuition about the search process; and \textit{(ii)} models that are expressed in more formal (mathematical) language~\cite{azzopardi2011economics,fuhr2008probability,wang2009portfolio,carterette2011effectiveness_evaluation,baskaya2013model,thomas2014model}. The latter category of model provides \textit{predictive power} about why users behave in a certain way. As such, they can be used as the basis of \textit{simulations of interaction}~\cite{azzopardi2010simint}. Here, a model of searcher behaviour that provides a credible approximation of reality can be used to ground simulations to examine what may happen under under given circumstances. 

Despite the advantages that simulations provide, formulating such descriptive models is non-trivial. Contemporary SERPs for example are complex user interfaces, with new components (e.g., \textit{entity cards}~\cite{navalpakkam2013measurement}) added all the time. In contrast, searcher models typically assume a simple SERP in the format of the traditional \textit{ten blue links}~\cite{hearst2009search}. Numerous studies have been undertaken on this more simplistic design, such as the cost of scrolling~\cite{azzopardi2016two, peytchev2006web, albers2002information}, typing~\cite{ong2018qwerty, crescenzi2015time} or response time lag~\cite{maxwell2014stuck, schurman2009performance}.

\paragraph{\textbf{Subtopics}} The \textit{Information Retrieval (IR)} community primarily considers the notion of subtopics from a system-centred point of view, with prior works focusing on ranking functions optimised for subtopic retrieval and result diversification~\cite{zhai2015beyond,jiang2017learning,dai2005minimal,nguyen2014leveraging,zuccon2010estimating}. Automatic subtopic (structure) extraction has also been investigated, generally based on a given starting query or document~\cite{takaki2004associative,hearst1993subtopic}. The influence of subtopic characteristics on users has not been frequented in IIR. One exception is by C\^amara et al.~\cite{camara2021searching}, who provided study participants with a list of subtopics and (visual) indicators about the extent of their subtopic exploration. The impact of subtopic ordering on users was not investigated. 

\paragraph{\textbf{\sal{}}} We ground our work in the domain of learning which has attracted considerable attention in recent years. Beyond studies investigating how learning-oriented searches are conducted~\cite{eickhoff2014lessons,moraes2018contrasting} and how to measure learning occurring in search sessions~\cite{bhattacharya2018relating,eickhoff2014lessons,wilson2013comparison}, multiple recent studies have investigated the impact of certain user characteristics and user actions on learning during search sessions---examples include the impact of domain knowledge~\cite{o2020role,wildemuth2004effects}, source selection strategies~\cite{liu2018information}, and the cognitive abilities of users~\cite{pardi2020role}. While observational studies are numerous, works proposing novel retrieval algorithms~\cite{syed2017retrieval} and novel interface elements~\cite{camara2021searching,roy2021note} to support learning whilst searching remain sparse. 

\section{Subtopic-Aware Complex Searcher Model (\texttt{SACSM})}\label{sec:model}

For our study, we augment the \csm{}~\cite{maxwell2015stopping, maxwell2015initial} to be subtopic-aware, turning it into the \sacsm{}. The \csm{} is a conceptual model of the IIR process (or a \textit{search session}), describing the flow of activities and decisions that a searcher undertakes when interacting with a search engine.
The \csm{} is built on the work of other conceptual models of the IIR process, such as the models of Baskaya et al.~\cite{baskaya2013model} and Thomas et al.~\cite{thomas2014model}. Conceptual models provide us with the necessary scaffolding from which we can expand and develop the model further for a SAL context---and instantiate the model in such a way that we can run our simulations of interaction~\cite{azzopardi2010simint}.

\begin{figure*}[ht!]
\includegraphics[width=1.0\textwidth]{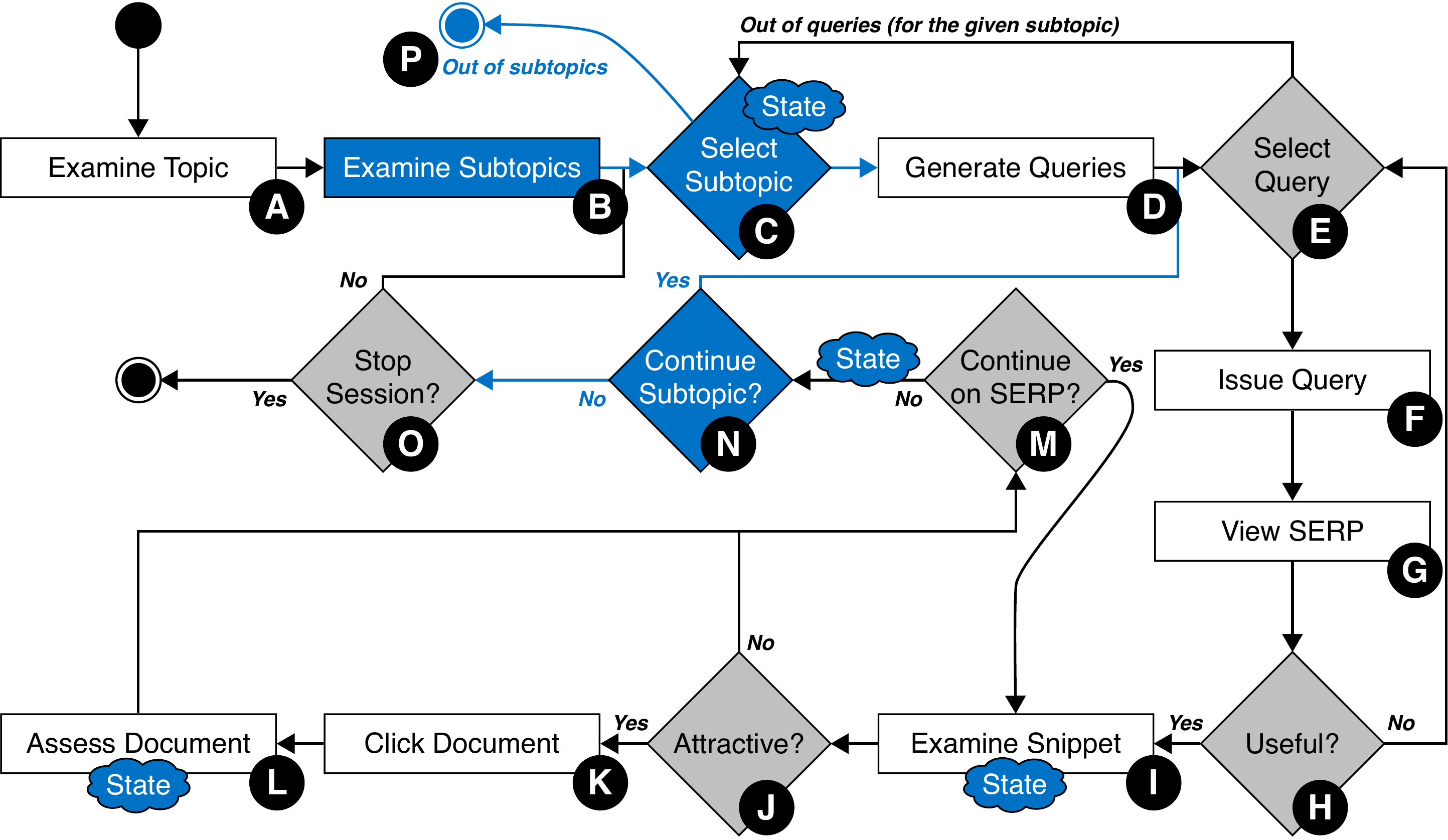}
\centering
\caption{The \textit{Subtopic-Aware Complex Searcher Model (\sacsm{})}. Changes from the \csm{} are highlighted in {\textbf{\color{modelblue}blue}}. Refer to \S\ref{sec:model} for more about the sequence and shapes.}
\vspace*{-5mm}
\label{fig:csm_sal}
\end{figure*}

The \sacsm{} is illustrated in Fig.~\ref{fig:csm_sal}; it includes a series of additional activities and decision points (compared to \csm{}) pertaining to the idea of \textbf{subtopic selection}, with novel components highlighted in {\color{modelblue}\textbf{blue}}. Key activities are represented as boxes~\includegraphics[height=\fontcharht\font`\B]{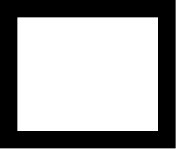}, with key decision points undertaken by subscribing agents represented as
 diamonds~\includegraphics[height=\fontcharht\font`\B]{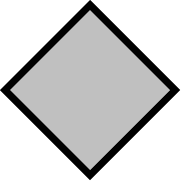}. Upon starting at~\includegraphics[height=\fontcharht\font`\B]{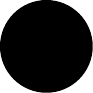}, a user (or, in the case of a simulation, an agent) following the \sacsm{} will first examine the given \textbf{topic}~\circled{A}. \sacsm{} then directs the agent to examine a list of the provided \textbf{subtopics}~\circled{B} for the given topic, before then deciding \textbf{what subtopic}~\circled{C} to examine in detail. From here, the agent will \textbf{consider a number of potential queries}~\circled{D} to issue pertaining to the selected subtopic, before \textbf{selecting a query}~\circled{E} to issue~\circled{F}. The agent will then obtain an `overview' of the SERP~\circled{G}, and decide whether to \textbf{enter it}~\cite{maxwell2018information}~\circled{H}---and if they do, they begin to \textbf{examine a snippet}~\circled{I}. If the present snippet is \textbf{sufficiently attractive}~\circled{J}, the agent will \textbf{click the associated link}~\circled{K}, and \textbf{assess the document}~\circled{L} for usefulness and/or relevancy, before deciding to \textbf{continue on the SERP}~\circled{M} (and examining further snippets if so). If not, the decision to \textbf{continue with the current subtopic}~\circled{N} is then made. If this is the case, further queries are issued~\circled{E}---meaning that the snippet and document examination activities are repeated for the results of the new query. This also means that subtopic exploration can entail multiple queries. If the agent instead decides to \textbf{abandon the subtopic}~\circled{N}, they must then decide whether to \textbf{stop the search session}~\circled{O} altogether. This process is repeated until all subtopics have been exhausted by the agent~\circled{P}, or some other condition is met---such as running out of session time.
 
Note that compared to previous instantiations of the \csm{}~\cite{maxwell2015stopping, maxwell2015initial, maxwell2016simulating, maxwell2018information}, we have removed activities and decision points about assessing documents for relevance. Unlike simple search sessions, with \emph{atomic} information needs, a SAL task generally has a more complex and nuanced need~\cite{eickhoff2014lessons}. Therefore, we are interested in examining the \textit{content} of retrieved documents (and thus learning from them)---not simply whether the documents themselves are considered relevant, as has been the norm for prior simulations of interaction~\cite{maxwell2019thesis}.

In order to keep track of terms/concepts that are examined by agents subscribing to the \sacsm{} (as vocabulary learning is a typical manner to measure learning gains in SAL~\cite{syed2018exploring,roy2020exploring,camara2021searching}), we must also incorporate some type of \textit{state} within it. This state model was considered in the study by Maxwell and Azzopardi~\cite[Fig.~3]{maxwell2016simulating} through the \textit{User State Model (USM)}, which \textit{``represents the user's cognitive state''}. Instead of representing the USM as a global, session-based model accumulating state and knowledge of the information examined, we consider a state model for the individual subtopics examined by agents. Each subtopic state consists of a representation of the terms observed by the agent to help them identify fundamental terms about the subtopic, which is used for query generation and determining what snippets (based on the snippet text provided by the underlying retrieval system) should be clicked on, with the corresponding document examined in more detail. Agents following this model only accrue knowledge when examining documents in full, deterministically deducing whether a document is worth examining without recourse to any relevance judgements. The state model is updated at points represented by~\includegraphics[height=\fontcharht\font`\B]{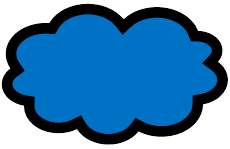} in Fig.~\ref{fig:csm_sal}. 
\section{Experimental Method}\label{sec:methods}
In this section, we describe the details of our instantiation of the \sacsm{} and our simulations. We start by defining how we instantiate each of the components of the \sacsm{} from Fig.~\ref{fig:csm_sal}, dividing them between fixed (i.e., no difference between agents) and variable components (i.e., changes between each agent). By tweaking the variable components, we can instantiate agents that simulate users with different characteristics. For example, an agent with a high \ls{} \textit{(how fast am I at learning new terms?)}, low \ex{} \textit{(how much content should I explore?)} and low \tl{} \textit{(how liberal am I at clicking links?)} simulates a learner that can quickly absorb new concepts, while only skimming through documents and clicking on almost all documents presented to them. We also outline our search setup, our simulation setup, as well as the datasets, and topics (and subtopics!) used. 

\subsection{Fixed \sacsm{} Components}\label{sec:components}

We instantiate the \sacsm{} in various ways to evaluate how different \subtopic{} strategies performed for different types of users. Although the \sacsm{} has many activities and decision points to instantiate, we fixed a number of these to reduce the space that we were required to examine.

\paragraph{\textbf{Query Generation}} We use the \textbf{$QS3^+$} querying strategy proposed by Maxwell and Azzopardi~\cite{maxwell2016simulating}, where three query terms are selected from a language model learned from the documents the agent has already explored (plus the topic description). Previous user studies in the \sal{} domain~\cite{camara2021searching, roy2021note} have shown that three query terms per query is reasonable, and close to what real-world searchers use.

\paragraph{\textbf{SERP Examination}} Considered by Maxwell and Azzopardi~\cite{maxwell2018information}, SERP examination strategies provide users with the ability to survey a SERP before committing to examining it in detail. Here, We choose to reduce the complexity of our agents (and explored space) and use the \textit{Always Examine} approach---agents always enter the SERP and examine at least one result snippet.

\paragraph{\textbf{User Interaction Costs}} To realistically mimic how long agents should spend on each phase of their search process, we present in Table~\ref{tab:costs} the costs (in seconds) from the interaction data of a prior user study~\cite{camara2021searching}. Note the high document examination cost---as participants of the user study were attempting to formulate ideas about concepts, they spent on average longer on documents when compared to other, non-SAL based studies (e.g., ~\cite{maxwell2018information}). We also note that the total session times influence the stopping behaviours of agents since, when agents reach the time limit of their sessions, they automatically stop---regardless of the number of remaining queries to be issued, as generated by the \textbf{$QS3^+$} strategy.

\paragraph{\textbf{Snippet-Level Stopping Strategies}} Different \textit{snippet-level stopping strategies} can be employed, generally classified between \textit{fixed} (i.e., the agent will evaluate snippets until a certain depth) or \textit{adaptive} strategies (i.e., the number of snippets evaluated may change depending on factors like agent state, presented snippet content, etc.). For our agents, we use only a fixed snippet-level stopping strategy, where agents examine snippets to a depth of $10$. This is a reasonable depth to examine to, and avoids issues with SERP pagination.

\begin{table}[t!]
\caption{Interaction costs grounding our agents, as derived from C{\^a}mara et al.~\cite{camara2021searching}.}
\centering
\begin{tabular*}{1\textwidth}{l @{\extracolsep{\fill}} c}
\toprule
\multicolumn{1}{l}{\textbf{Time required to...}} & \multicolumn{1}{l}{\textbf{Value (in seconds)}} \\
\midrule
...issue a query  & 9.42 \\
...examine a SERP  & 2.00 \\
...examine a result snippet  & 3.00 \\
...examine a document  & 80.00 \\ \midrule
\textbf{Total session time}  & 2400 \\ \bottomrule
\end{tabular*}
\label{tab:costs}
\vspace*{-.25cm}

\end{table}

\subsection{Variable \sacsm{} Components}\label{sec:variable}
\begin{figure*}[t!]
\includegraphics[width=\textwidth]{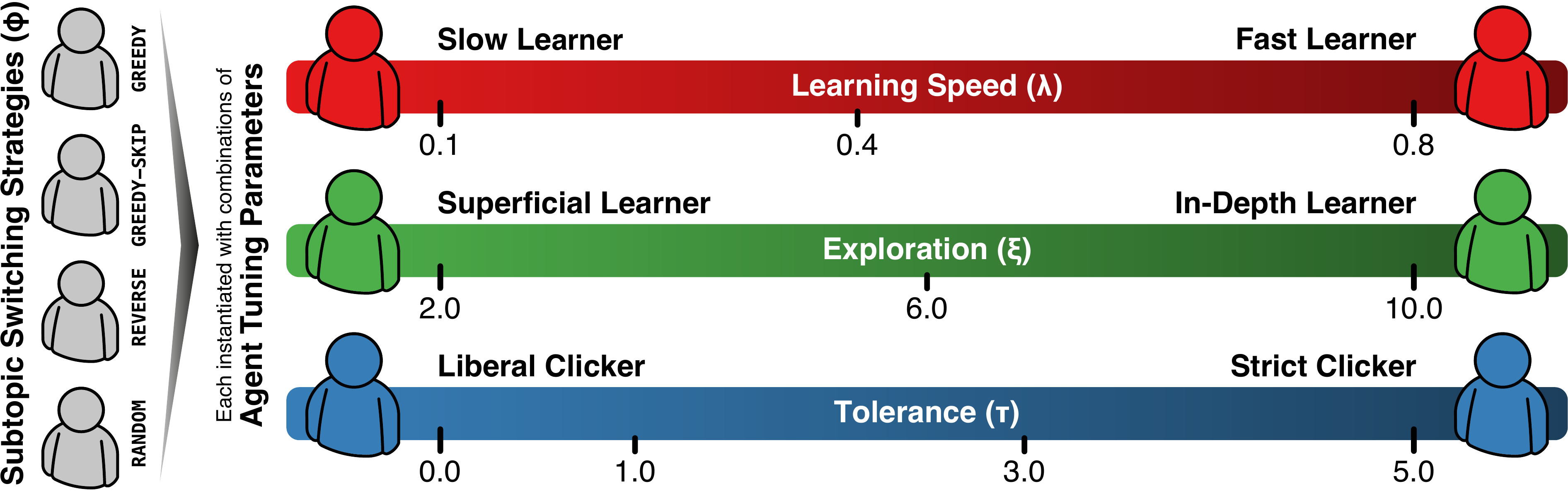}
\centering
\caption{Overview of the four variable parameters for instantiating simulated agents.}\vspace*{-5mm}
\label{fig:learner-spectrum}
\end{figure*}

For this study, agents can be instantiated using four variables, according to the type of user to be simulated. An overview of these variables is presented in Fig.~\ref{fig:learner-spectrum}.

\paragraph{\Subtopic{} (\sub{})} We propose four different strategies for agents to select and switch between subtopics during their search sessions. These implement the \textbf{Select Subtopic} decision point, as shown in Fig.~\ref{fig:csm_sal}. To determine whether agents have explored a subtopic sufficiently, we use a method similar to the approach outlined by C{\^a}mara et al.~\cite{camara2021searching} for tracking subtopic exploration. Each clicked document is embedded using \texttt{SBERT}~\cite{reimers2019sentence} and compared---using the dot product---to pre-computed embeddings for each subtopic of the current topic, as extracted from their \textit{Wikipedia} articles\footnote{Refer to \S\ref{sec:method:simsetup} for more information on the use of Wikipedia articles.}. Therefore, each document clicked by an agent will update an internal state tracker for each subtopic, summing how much the agent \textit{`explored'} each subtopic. We evaluate four strategies.

\begin{itemize}
    \item{\greedy{}  For this strategy, an agent examines each subtopic in turn, according to the order provided by the respective Wikipedia article, only deciding to move to the next subtopic when they have achieved a certain level of progress. Intuitively, this would be the most rational type of user, since they follow a subtopic ordering that is optimised for human understanding (i.e., the order comes from a Wikipedia page). In other words, they will attempt to master one subtopic before moving to the next (prescribed) topic.}
    
    \item{\greedys{} Instead of the above, an agent subscribing to \greedys{} moves to the next subtopic with the \textit{next lowest completion value.} This instantiated agent attempts to minimise the number of documents to be read by querying in a domain with lesser knowledge.}
    
    \item{\reverse{} This strategy is similar to \greedy{}, but the agent examines the subtopics in reverse order as presented from the corresponding Wikipedia article. The rationale here is that an agent attempts to game the system by first learning the most complex subtopics \textit{before} moving to easier ones.}
    
    \item{\random{} This strategy randomly selects a new subtopic after each query, with no predefined order. This strategy models a non-rational learner, and serves as a lower bound for our experiments.}
\end{itemize}

\paragraph{\Lam{}} (\ls{}) This parameter is the same $\lambda$ from the language model proposed by Maxwell and Azzopardi~\cite{maxwell2016simulating}. 
It controls how much an agent relies on their acquired knowledge (i.e., novel terms) when considering if a given snippet should be clicked or not. The language model is updated every time the agent clicks on a relevant snippet. In addition, a \textit{Maximum Likelihood Estimator}~\cite{meij2009query} is used for deciding if a given snippet is attractive or not. An agent with a low \ls{} gives lower weights to terms learned during the session, simulating a \textit{slow learner}. An agent with a higher \ls{}, in turn, mimics a user that quickly incorporate new terms, being a \textit{fast learner}. In our simulations, we use \ls{}~$\in[0.1, 0.4, 0.8]$.
\vspace{-4mm}
\paragraph{\Limit{}} (\textbf{\ex{}}) This parameter controls how much the agent should explore a subtopic before being satisfied by what it has \textit{`learnt'}. A lower number implies that such an agent that is only `skimming' trough the topic, inspecting only a few documents per subtopic. In contrast, a higher value implies that such an agent is willing to explore deeper within each subtopic. For the simulations reported in this paper, we trial \ex~$\in[2.0, 6.0, 10.0]$.
\vspace{-4mm}
\paragraph{\Threshold{}} (\tl{}) Finally, this parameter is the threshold that controls how attractive a snippet should be to be clicked~\cite{meij2009query}. An agent with low \tl{} is a \textit{strict clicker}, clicking on fewer, \textit{`safer'} snippets, while a higher \tl{} implies a \textit{liberal clicker} agent, more willing to explore. In our simulations, we trial \tl~$\in[0.0, 1.0, 3.0, 5.0]$.

\subsection{Simulation Setup}\label{sec:method:simsetup}
The setup of our experiments follows that of the user study presented by Câmara et al.~\cite{camara2021searching}. 
Here, we use the same eight topics extracted from the \car~2017 dataset~\cite{TRECCAR2017}, as shown in Table~\ref{tab:keywords}. Subtopics were also derived from the \car~dataset: they were extracted from first-level headings of the respective \textit{Wikipedia} articles as they were in December 2017---the dataset's creation.

 \begin{table}[t!]
\vspace{-0.5cm}
\centering
\caption{\textbf{\#} of subtopics and distinct keywords \textbf{(KW)} for each topic. For each subtopic, we determine the ten KWs with higher TF-IDF on the respective subtopic section on Wikipedia. A KW may appear in the top ranks of several subtopics. KW difficulty is given by the age-of-acquisition, as proposed by Kuperman et al.~\protect\cite{kuperman2012age}.}
\scriptsize
\resizebox{\columnwidth}{!}{%
\begin{tabular}{l @{\extracolsep{\fill}} ccc}
\toprule
\textbf{Topic}                             & \textbf{\#Subtopics} & \textbf{\#Unique KWs} & \textbf{KW Difficulty}\\ \midrule
Ethics                            & 6               & 49                    & 10.85                      \\
Genetically Modified Organism     & 5               & 33                    & 9.97                       \\
Noise-Induced Hearing Loss        & 8               & 56                    & 8.85                       \\
Subprime Mortgage Crisis          & 8               & 52                    & 9.81                       \\
Radiocarbon Dating  & 4               & 35                    & 9.77                       \\
Business Cycle                    & 4               & 32                    & 10.70                      \\
Irritable Bowel Syndrome          & 10              & 72                    & 9.88                       \\
Theory of Mind                    & 8               & 67                    & 9.63                       \\ \bottomrule
\end{tabular}
}
\label{tab:keywords}
\vspace{-0.5cm}
\end{table}

Our study uses the \textit{Bing Search API} to provide a ranking for queries issued by real-world users and our simulated agents. We used a manually curated blocklist\footnote{\url{https://github.com/ArthurCamara/CHIIR21-SAL-Scaffolding/blob/master/data/blocklist.txt} (All URLs last accessed January 18\textsuperscript{th}, 2022.)} of URLs serving \textit{Wiki}-style clones to filter results returned from the Bing API to prevent agents from encountering a single page that would give them all the information on all subtopics at once. This encourages agents to examine multiple documents and issue multiple queries to find information pertinent to their learning task. To match with our stopping strategy (see \S\ref{sec:components}), 10 results per page were presented to agents.

\vspace{-0.5mm}
\section{Results}\label{sec:results}
\vspace*{-2mm}
By combining all values of \ex{}, \ls{}, \tl{} and \sub{}, we instantiate $144$ unique agents (using a modified version \simiir{}~\cite{maxwell2016simulating}), and run each agent over all of the topics shown in Table~\ref{tab:keywords}. Our version of \simiir{}---together with the raw outputs of our simulations---are available at \url{https://github.com/ArthurCamara/simiir_subtopics/}. With some methods being non-deterministic, each agent was run ten times---with the average reported. In total, we ran a total of $11,520$ simulations. 
We show representative examples for each set of measures. In all plots, the $x$ axes denote how many documents the agent examined during a search session. While values on $y$ axes may seem low, they are averaged over a large number of simulations with varying degrees of complexity.


 
\begin{table}[t!]

\caption{Overview of (average) measures across agents and subtopic switching strategies, and real learners extracted from the \progress{} cohort from C\^{a}mara et al.~\cite{camara2021searching}.}

\centering
\begin{tabular*}{1\textwidth}{l @{\extracolsep{\fill}} ccc}
\toprule
    \multicolumn{1}{l}{\scriptsize{\textbf{Strategy (\sub{})}}} & \multicolumn{1}{l}{\scriptsize{\textbf{\#Queries Issued}}} & \multicolumn{1}{l}{\scriptsize{\textbf{\#Snippets Examined}}} & \multicolumn{1}{l}{\scriptsize{\textbf{\#Documents Clicked}}} \\
\midrule
\progress{}(N=36) & 11.86($\pm7.60$) & 152.44($\pm 84.23$) & 18.50($\pm 9.56$) \\
\midrule
\greedy{}  & 13.05($\pm 14.93$)           & 133.37($\pm176.29$)          & 21.32($\pm7.60$)          \\
\greedys{} & 13.05($\pm 15.13$)           & 133.23($\pm 177.26$)           & 21.44($\pm 8.88$)       \\
\reverse{} & 12.01($\pm 14.55$)           & 123.28($\pm 173.34$)           & 21.42($\pm 8.78$)      \\
\random{}  & 13.03($\pm 16.07$)           & 117.61($\pm 155.80$)           & 21.82($\pm 8.30$)      \\ \bottomrule
\end{tabular*}

\label{tab:metrics}

\end{table}

Table~\ref{tab:metrics} shows the average value for key measures over all agents of each \sub{}, over the eight topics. In the first row, we also show the measures from the \progress{} cohort from C\^{a}mara et al.~\cite{camara2021searching}. Our simulated agents are similar on these measures compared to how real-world learners would behave, with a similar number of queries, snippets examined, and documents clicked. While a high deviation is expected, recall that this is an average of $288$ agents with a large variation on their parameters (compared to only $36$ real-world learners). Results show that our agents are indeed similar to real-world learners. To address \textbf{RQ}, we break down our analysis further into three sub-questions.

\paragraph{\textbf{How many keywords can the agents find?}}\label{sec:keywords}
\begin{figure}[t!]
    \centering
    \includegraphics[width=\textwidth]{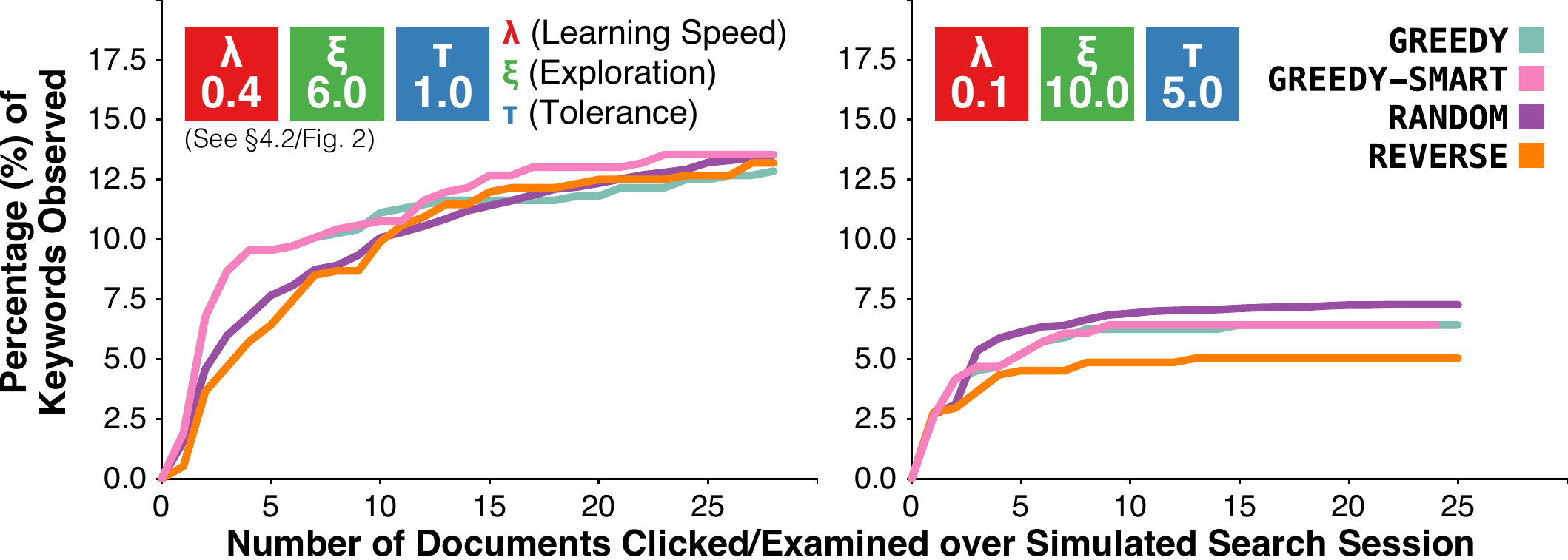}
    \caption{Accumulated percentage of the keywords seen for two different agents (averaged over all topics, weighted by number of keywords) with varying \ex{},~\ls{} and~\tl{}.}\vspace*{-5mm}
    \label{fig:keywords}
\end{figure}

To measure how well the agents can find documents with a high concentration of potentially valuable keywords\footnote{As noted in \S\ref{sec:model}, we do not have explicit relevance judgements.}, we extracted ten keywords for each subtopic from their respective paragraphs from the topic’s Wikipedia article.\footnote{As an example, the following are extracted keywords for the topic \textit{Ethics}: \texttt{ethical, ontology, propositions, consequentialism, normative} and \texttt{principles}.} To do this, we begin by ranking all terms from their portions of the articles (excluding stopwords) by their TF-IDF, with the IDF computed over
the whole TREC CAR Wikipedia dump---and selecting the top 10 terms as keywords. We use this subtopic-wise approach (instead of extracting keywords from the whole article) to ensure a fair distribution of keywords over all subtopics, providing a less biased overview of how the agent is performing over the topic. Therefore, each topic has a different number of keywords, reflected by the number of subtopics it contains. Table~\ref{tab:keywords} shows how many subtopics and unique keywords each topic contains. This setup is similar to previous~\sal{} user studies~\cite{camara2021searching, roy2020exploring, roy2021note}, where study participants were asked to define a list of concepts before and after their search session to evaluate their knowledge gain. We can mimic this setup throughout the entirety of an agent's search session by requiring the keyword to appear at least a few times (in our case, five) in the documents \textit{`read'} by the agent.

For two agents, Fig.~\ref{fig:keywords} shows how many keywords each approach for \sub{} discovers during their search sessions after reading a certain number of documents. At the beginning of the search sessions, we observed that agents instantiated with~\greedy{} and~\greedys{} strategies found keywords faster than agents with~\random{} or~\reverse{}. However, this difference diminishes over time. This is expected, since the subtopics ordering comes from Wikipedia articles, which are optimised for human understanding. Therefore, an agent that searches for subtopics in order has a higher probability of encountering documents with a higher number of keywords earlier in the session when compared with one that does not. We can also note that~\random{} with higher~\tl{} found more unique keywords in total, given their high probability of clicking in any document.

\vspace*{-3mm}
\paragraph{\textbf{Are the agents exploring enough of the subtopics?}}\label{sec:subtopics}
\begin{figure}[t!]
    \centering
    \includegraphics[width=\textwidth]{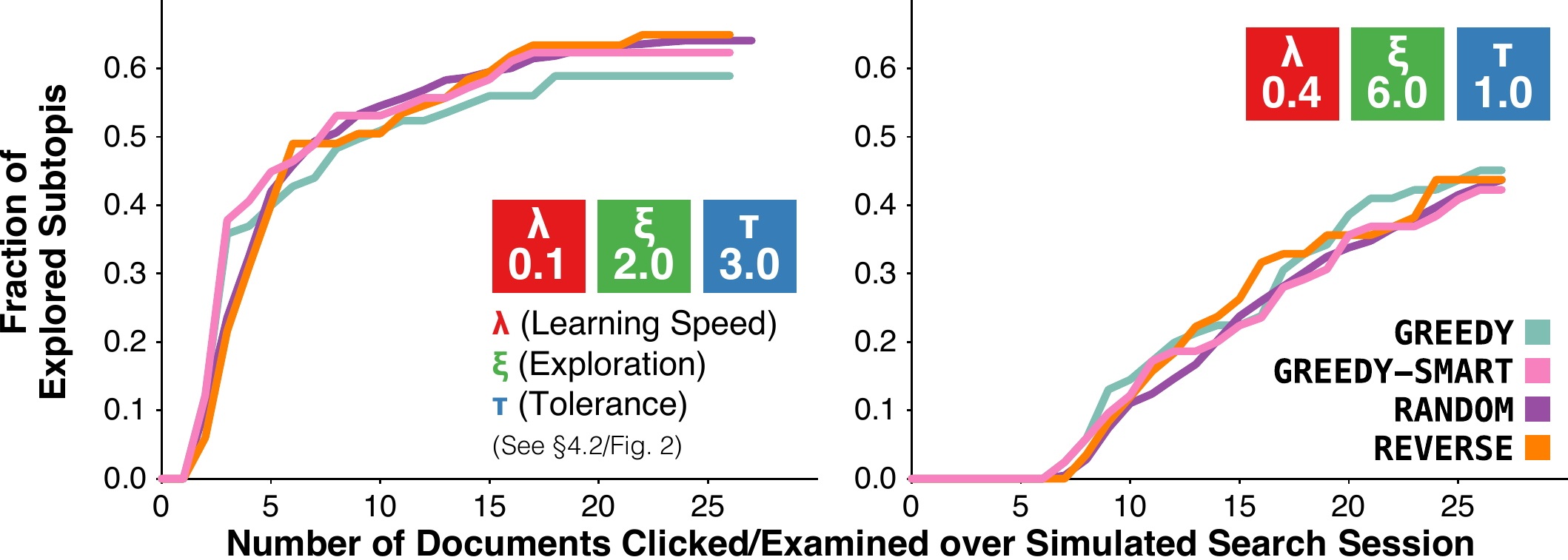}
    \caption{Fraction of fully explored (i.e agent reached~\ex{} value) for two different agents (averaged over all topics, weighted by number of keywords) with varying \ex{},~\ls{} and~\tl{}.}
    \label{fig:subtopics}
    \vspace{-0.5cm}
\end{figure}

Another way to measure how the agents behave is by investigating how their internal \emph{Subtopic Trackers} evolve during the session (as explained in~\S\ref{sec:model}). If an agent can reach~\ex{} for a given subtopic in a few documents, we can infer that they could quickly find documents related to that subtopic. Fig.~\ref{fig:subtopics} shows a similar trend to that observed previously, with agents using~\greedy{} and~\greedys{} strategies clicking on documents that advance their internal tracking faster. This implies that these strategies effectively lead the agents towards better documents faster. 

\vspace*{-3mm}
\paragraph{\textbf{Are the agents following the order of the subtopics?}}\label{sec:ordering}

\begin{figure}[t!]
    \centering
    \includegraphics[width=\textwidth]{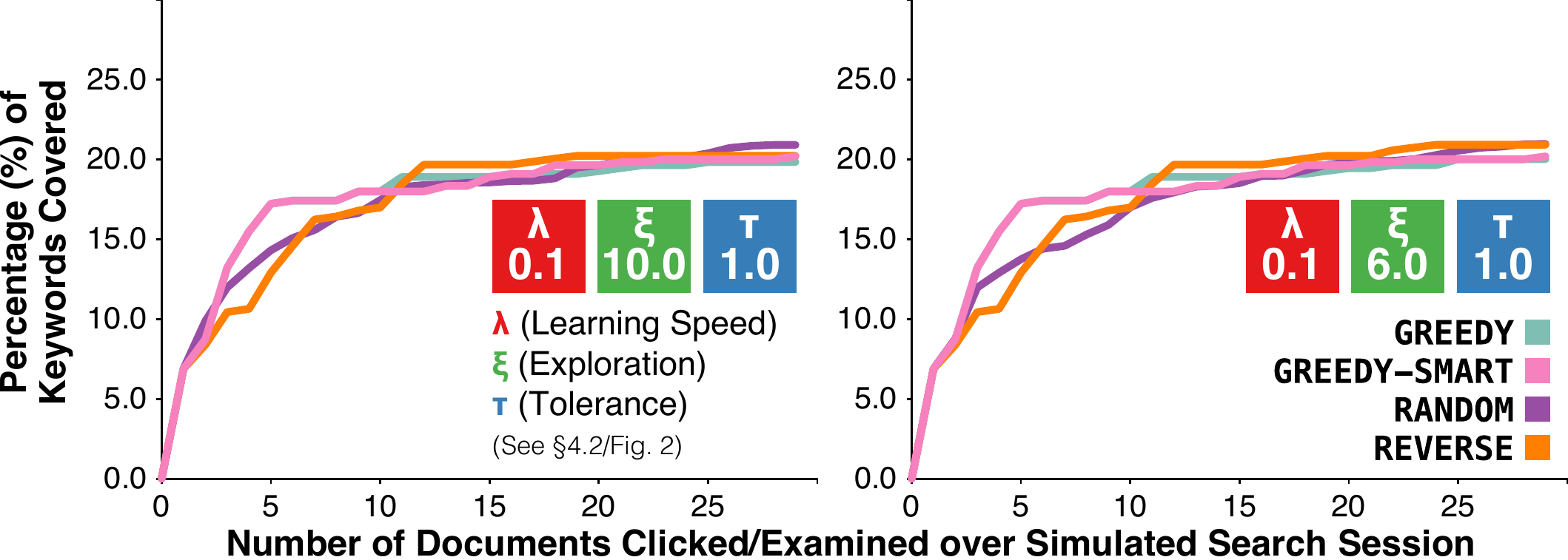}
    \caption{Fraction of keywords properly \textit{`learned'} by two different agents (averaged over all topics, weighted by number of keywords) with varying \ex{},~\ls{} and~\tl{}.}
    \vspace{-0.5cm}
    \label{fig:ordering}
\end{figure}

While the previous measures show that the agents are indeed effective in finding documents related to the topic, they fail to incorporate another essential learning feature, namely that keywords have dependencies between them. We assume that, for an agent to comprehend what a keyword means entirely, they have to comprehend at least some other, more basic concepts related to the topic at hand. Therefore, an agent that can find documents so that they will encounter more primary keywords for the topic (i.e., that appear earlier in the Wikipedia article) earlier in the session before facing more complex keywords (i.e., that appear later in the Wikipedia article) is more desirable for a \sal{} environment. As an example, consider the keyword \texttt{consequentialist} for the topic \textit{Ethics}. Before an agent can adequately understand what it means in this context, they probably need to understand other concepts, like \textit{virtue} and \textit{morality}. Therefore, for this analysis, we consider a keyword to be \textit{`learned'} after the agent has already encountered a certain number of keywords that appear prior to it in the original Wikipedia article. To account for possible noises in our keyword extraction method, we define this number as 50\% of the keywords seen prior to the current one (e.g., the keyword \texttt{consequentialist} is the $19$\textsuperscript{th} out of $49$ keywords to appear in the list of extracted terms for the topic \textit{Ethics}).
Therefore, we only consider a given keyword as learned after the agent has learned at least 30\% of the prior terms.\footnote{This number was decided experimentally, as it showed to be the best to distinguish between the different methods trialled.} As seen in Fig.~\ref{fig:ordering}, we see similar behaviour to the one observed above, with \greedy{} and \greedys{} outperforming \random{} and \reverse{}, with the difference slowly disappearing throughout the search session. Again, almost all agents repeat this behaviour.
These results show that our simulations are close to real users and that there is a clear difference between strategies, with \greedy{} and \greedys{} following the logical structure of the subtopics, and generally being better strategies for agents exploring the subtopic space. Consequently, these should be taken into account when simulating agents for \sal{} scenarios.

\vspace{-0.5cm}
\section{Conclusions}\label{sec:conclusions}
\vspace*{-2mm}
We have proposed a novel user model for simulating agents focused on \sal{} tasks: the \emph{Subtopic-Aware Complex Searcher Model}, \sacsm{}. Recalling our original research question which considered how different subtopic switching strategies (\sub{}) affected the behaviour or simulated agents, we show that strategies that mimic a rational user (i.e., \greedy{} and \greedys{}) are more effective at \emph{finding keywords}, \emph{exploring subtopics} and \emph{following subtopic structure} when compared to other strategies. With $1,520$ simulations, our study is the first (to the best of our knowledge) that focuses on simulated agents for Search as Learning, enabling future works in both \sal{} and IIR that may require large quantities of user data, such as Reinforcement Learning models and studies on how changes in the search system may impact the behaviour of learners. To further help research efforts, we also make public our implementation of the \sacsm{}, built on top of the already established \simiir{} framework.
%


\bibliographystyle{splncs04}
\bibliography{references}

\end{document}